\documentclass[11pt,aps,floatfix,tightenlines]{revtex4}
\newcommand{\diff}{{\rm\,d}}                    
\def\p{\mbox{\boldmath $p$}}
\usepackage{epsfig}
\usepackage{graphicx}
\usepackage{graphics}
\usepackage{xspace}
\usepackage{amssymb}
\usepackage{amsmath}
\usepackage{latexsym}
\usepackage{natbib}
\usepackage{mathrsfs}
\usepackage{url}
\def\q{\mbox{\boldmath $q$}}

\begin{document}

\title{Nuclear Physics with Electroweak Probes}

\author{Carlotta Giusti}

\affiliation{Dipartimento di Fisica Nucleare e Teorica, Universit\`{a} degli 
Studi,\\ 
and INFN, Sezione di Pavia, via Bassi 6 I-27100 Pavia, Italy}

\begin{abstract}
The research activitities carried out in Italy during the last two years in 
the field of theoretical nuclear physics with electroweak probes are 
reviewed. Different models for electron-nucleus and neutrino-nucleus scattering 
are compared. The results obtained for electromagnetic reactions on few-nucleon 
systems and on complex nuclei are discussed. The recent developments in the 
study of electron- and photon-induced reactions with one and two-nucleon 
emission are presented.

\end{abstract}

\maketitle

\section{Introduction}

Several decades of theoretical and experimental investigation in the field of 
nuclear physics with electromagnetic probes have yielded a wealth of information
on nuclear structure and interaction mechanisms \cite{book} . An important contribution has 
been given by the Italian theory groups that since many years have been working 
in the field, within many international collaborations and in close connection
with the experimental activities.

In spite of so many years of investigation and of the great progress achieved,
some interesting aspects are still unclear and
there is yet much to be learned. In the last few years theories have improved
remarkably but some data are old, incomplete or not accurate 
enough to disentangle interesting effects. The need of a more
complete experimental program to investigate nuclei with electromagnetic probes
was already pointed out by G. Co' in the previous report two years ago
\cite{CoCor} . 

In recent years almost all the Italian groups working in electron scattering 
have applied their models to $\nu$-nucleus scattering. Although the two 
situations are different, the extension to neutrino scattering of the electron 
scattering formalism is straightforward. Moreover, electron scattering is the 
best available guide to determine the prediction power of a nuclear model. 
 
The observation of neutrino oscillations and the proposal and realization of
new experiments, aimed at determining neutrino properties with high accuracy,
renewed interest in neutrino scattering on complex nuclei. In fact, neutrinos 
are elusive particles. They are chargeless, almost massless, and only weakly 
interacting. Their presence can only be inferred detecting the particles they 
create when colliding or interacting with matter. Nuclei are often used as 
neutrino detectors providing relatively large cross sections. 
The interest in $\nu$-nucleus scattering extends to different fields, such as 
astrophysics, cosmology, particle and nuclear physics. In hadronic and nuclear 
physics neutrinos can excite nuclear modes unaccessible in electron scattering, 
can give information on the hadronic weak current and on the strange nucleon
form factors  \cite{MeucciCor} . 
Thus, neutrino physics is of great interest and involves many different
phenomena. The interpretation of data, however, requires a detailed knowledge 
of the $\nu$-nucleus interaction as well as reliable cross section 
calculations where nuclear effects are properly taken into account. 

The different models used to treat electron- and $\nu$-nucleus scattering are
discussed  in Sec. 2. The treatment of the two-body weak axial
current is presented in Sec. 3. Scaling and superscaling in lepton-nucleus
scattering are considered in Sec. 4. Electromagnetic reactions in few-body
and complex nuclei are reviewed in Sects. 5 and 6, respectively.

\section{Electron-Nucleus and Neutrino-Nucleus Scattering}

Lepton-nucleus scattering is usually described in the one-boson exchange
approximation. The exchanged boson is the photon in the case of the
electromagnetic interaction and the $Z^0$ or $W^\pm$ in the case of the weak 
interaction. In electron scattering the invariant amplitude is given by
the sum of the one-photon and the one-$Z^0$ boson exchange term. The first term
is parity conserving and the second one is Parity Violating (PV). PV
Electron Scattering (PVES) requires a polarized incident electron and is 
interesting to study the strange nucleon form factors \cite{MeucciCor} . 
In $\nu$ ($\bar\nu$)-nucleus scattering the boson exchanged is the $Z^0$ for 
neutral-current (NC) scattering, i.e., ($\nu,\nu'$) [($\bar\nu,\bar\nu'$)], and 
the $W^+$ ($W^-$) for charged-current (CC) scattering, where a charged 
lepton is obtained in the final state, i.e., ($\nu,l^-$) [($\bar\nu,l^+$)]. 

Different processes can thus be considered. In any case, in the one-boson 
exchange approximation, the cross section is given in the form
\begin{equation}
\diff \sigma  = K L^{\mu\nu}\ W_{\mu\nu},
\label{aba:cs}
\end{equation}
where $K$ is a kinematical factor, the lepton tensor $L^{\mu\nu}$ 
depends only on the lepton kinematics, and  the nuclear response is contained 
in the hadron tensor $W^{\mu\nu}$, whose components are given by products of 
the matrix elements of the nuclear current  $J^{\mu}$ between the initial 
and final nuclear states, i.e.,
\begin{equation}
W_{\mu\nu} = \sum_f \, \langle \Psi_f\mid J^{\mu}(\q) \mid \Psi_i\rangle \, 
\langle \Psi_i \mid J^{\nu\dagger}(\q)\mid \Psi_f\rangle \, 
\delta(E_i+\omega-E_f),
\label{aba:wmn}
\end{equation}
where $\omega$ and $\q$ are the energy and momentum transfer, respectively.
Similar models are used to calculate $W^{\mu\nu}$ in electron- and 
$\nu$-nucleus scattering. 

In a schematic representation of the nuclear response to the electroweak probe,
different regions can  be identified. At low energy transfer, below the 
threshold for the emission of a nucleon from the target, the response is
dominated by discrete states, that can be treated in large model spaces. 
Above the continuum threshold there are giant resonance levels, collective
excitations that can be described within the Random Phase Approximation (RPA). 
Then, a large broad peak occurs at about $\omega=q^2/2m_{\mathrm N}$. 
Its position corresponds to the elastic peak in electron scattering by a free 
nucleon. In the region of the quasielastic (QE) peak the response is dominated 
by the single-particle (s.p.) dynamics and by 
one-nucleon knockout processes, where the interaction occurs on a quasifree 
nucleon which is emitted from the nucleus with a direct one-step mechanism. 
At higher energies mesons and nucleon resonances can be produced. 
For instance, at an excitation energy of $\sim$ 300 MeV the $\Delta$-peak 
corresponds to the first nucleon excitation.

Schematically, the response is the same for electron and neutrino scattering. 
However, the cross sections calculated for the (e,e$'$), ($\nu,\nu'$),
($\nu$,e$^-$), and ($\bar\nu$,e$^+$) reactions on O$^{16}$ in the same 
kinematic conditions within the continuum RPA have a quite different behavior
\cite{CoCor,CoB} . The difference is due to the different current in the 
electromagnetic and weak interactions: the neutrino cross sections
are dominated by the axial vector term that does not contribute
to electron scattering. As a consequence, it is necessary to be careful in
relying on the fact that a model able to give a good description of electron 
scattering will give a good description of neutrino scattering. In
spite of this warning, a model able to describe electron scattering data can 
be considered as a good basis to treat also $\nu$-nucleus  
scattering.

\subsection{Random Phase Approximation} 

Electron-nucleus and $\nu$-nucleus cross sections calculated within the
RPA are compared in Refs.~\cite{CoB,Co}. The RPA is an 
effective theory, aiming to describe the excitations of many-body systems, that 
has been widely and successfully applied in nuclear physics over a wide range 
of excitation energies.

The RPA describes the nuclear excited states as a linear combination of
particle-hole (ph) and hole-particle (hp) excitations. The combination
coefficients for each  state are obtained solving the secular RPA
equations that contain s.p. energies and wave functions as input. They are 
generated by a Woods-Saxon potential whose parameters have been fixed to 
reproduce the energies of the levels close to the Fermi surface and the rms 
radii \cite{CoB,Co} . The other input of the theory is the effective 
interaction, $V^{\mathrm{eff}}$, that is an effective interaction in 
the medium. Calculations are compared in Ref.~\cite{CoB,Co} with various 
zero-range and finite-range interactions, in order to point out the 
sensitivity of the RPA results to the choice of $V^{\mathrm{eff}}$. 

RPA calculations are necessary to produce giant resonances and collective
low-lying states. The results, however,  strongly depend on the effective
interaction used. Interactions equivalent from the spectroscopic point of view 
can produce different excited states. This indicates the limits of the RPA. In 
the evaluation of some observables, the use of effective interactions cannot 
substitute the explicit treatment of degrees of freedom beyond 1p-1h 
excitations. The use of quenching factors reduces the spreading of the results 
obtained with different interactions in the electron excitation and increases 
the spreading in the neutrino excitation. This is further evidence that 
electrons and neutrinos excite the same states in a different way. The 
uncertainty on the cross section is large for neutrinos of 20-40 MeV. These 
uncertainties have heavy consequences on the cross sections of low energy 
neutrinos. 
When the neutrino energy is above 50 MeV, the results are rather independent 
of the interaction and the inclusion of many-particle many-hole 
excitations reduces the RPA cross sections by a factor of $\simeq 10-15\%$.

\begin{figure}[ht]
\begin{center}
\includegraphics[height=90mm,width=70mm]{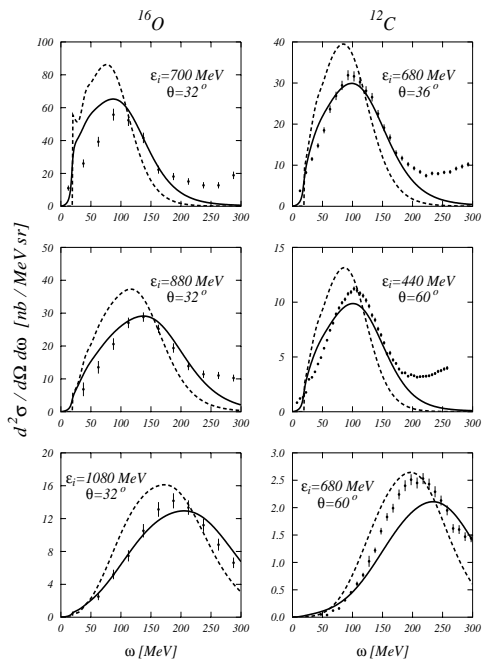}
\vspace{2mm}
\caption{ $^{16}$O(e,e$'$) and $^{12}$C(e,e$'$) cross sections in the QE 
region. The MF results are shown by the dashed lines. The inclusion of FSI 
produces the full lines. Data from Ref.~7. (From Ref.~5).}
\end{center}
\end{figure}
In the QE region RPA effects are rather small when a finite-range 
$V^{\mathrm{eff}}$ is used \cite{CoB,Co} . A zero-range interaction 
overestimates RPA effects. However, many-body effects beyond RPA are important, 
as it is shown in Fig. 1, where QE $^{16}$O(e,e$'$) and $^{12}$C(e,e$'$) cross 
sections calculated within the mean-field (MF) model, i.e., setting 
$V^{\mathrm{eff}}=0$, are compared to data. In the QE region these many-body 
effects, which in the RPA language are 
described as many-particle many-hole excitations, are usually called 
Final-State Interactions (FSI). Different treatments of FSI are available in 
the literature. In Fig. 1 \cite{CoB,Co} FSI are folded with a Lorentz 
function whose parameters are fixed by hadron scattering data \cite{Co88} and 
give a redistribution of the strength that is essential to reproduce (e,e$'$) 
data.

\subsection{Nuclear Effects and FSI in the Quasielastic Region}

In the QE region the nuclear response is dominated by one-nucleon knockout 
processes. 
A lot of theoretical and experimental work has been done to study the QE 
inclusive (e,e$'$) and exclusive (e,e$'$p) reactions \cite{book,BDS} .  
This work can be helpful to treat $\nu$-nucleus scattering.

In the QE $\nu$($\bar\nu$)-nucleus NC and CC scattering we assume that one 
nucleon is emitted
\begin{eqnarray}
\nu (\bar\nu) + \mathrm{A} & \rightarrow & {\nu'} (\bar\nu') +\mathrm{N} + 
( \mathrm{A} - 1)  \hspace{3cm} \mathrm{NC} \\
\nu (\bar\nu) + \mathrm{A} & \rightarrow & l^{-} (l^{+}) +
\mathrm{p}(\mathrm{n}) + ( \mathrm{A}-1) \hspace{2.3cm} \ \mathrm{CC} 
\end{eqnarray} 
In the NC scattering only the emitted nucleon can be detected. Thus, the cross 
section must be integrated over the energy and angle of the final lepton. Also 
the state of the residual (A-1)-nucleus is not determined and the cross 
section is summed over all the available states of the residual nucleus. The 
same situation occurs for the CC reaction if only the outgoing nucleon is 
detected. The cross sections are therefore semi-inclusive in the hadronic 
sector and inclusive in the leptonic one and can be treated as an (e,e$'$p) 
reaction where only the outgoing proton is detected. The exclusive CC process 
where, as in the case of (e,e$'$p), the charged final lepton is detected in 
coincidence with the emitted nucleon can be considered as well. 
The inclusive CC scattering where only the charged lepton is detected 
can be treated with the same nuclear models used for the inclusive (e,e$'$) 
scattering. 

QE $\nu$-nucleus scattering has been treated in 
Refs.~\cite{Meucci1,Meucci2,GMP} using the same relativistic models 
that were developed for the inclusive (e,e$'$) and the exclusive (e,e$'$p) 
reactions. These models include nuclear effects and FSI.

In the first order perturbation theory and in the Impulse Approximation (IA), 
the transition amplitude of the NC and CC processes where the outgoing nucleon 
is detected is described as the sum of terms similar to those appearing in the 
Relativistic Distorted Wave IA (RDWIA) for the (e,e$'$p) knockout 
reaction. The amplitudes for the transition to a specific state $n$ 
of the residual nucleus are obtained in a one-body representation 
\begin{equation}
\langle n;\chi^{(-)}_{\p_{\mathrm N}}\mid J^{\mu}(\q) \mid \Psi_i\rangle = 
\langle\chi^{(-)}_{\p_{\mathrm N}}\mid   j^{\mu}
(\q)\mid \varphi_n \rangle   \label{eq.amp}
\end{equation} 
and contain three ingredients: the one-body nuclear weak current $j^{\mu}$,
the one-nucleon overlap $\varphi_n = \langle n | \Psi_i\rangle$, that is a s.p. 
bound state wave function whose normalization gives the spectroscopic 
factor, and the s.p. scattering wave function $\chi^{(-)}$ for the outgoing 
nucleon, that is eigenfunction of a complex optical potential describing the 
FSI between the outgoing nucleon and the residual nucleus.

Bound and scattering states are calculated with the same phenomenological 
ingredients used for the (e,e$'$p) calculations. 
A pure SM description is assumed for the states $n$, i.e., $n$ is a one-hole 
state in the SM and a sum over all the occupied SM states is carried out. 
In these calculations FSI are described by a complex optical potential 
whose imaginary part gives an absorption that reduces the cross 
sections by $\sim 50\%$. The imaginary part accounts for the 
flux lost in a particular channel towards other channels.
This approach is conceptually correct for an exclusive reaction, where only 
one channel contributes, but it would be conceptually wrong for an inclusive 
reaction, where all the channels contribute and the total flux must be 
conserved. For the semi-inclusive process where an emitted nucleon is 
detected, some of the reaction channels which are responsible for the imaginary 
part of the optical potential are not included in the experimental cross 
section and, from this point of view,  it is correct to include the 
absorptive imaginary part of the potential. 

In the inclusive scattering FSI can be treated in the Green's Function 
Approach (GFA), that was firstly applied 
to the QE (e,e$'$) scattering in a nonrelativistic \cite{ee} and in a 
relativistic \cite{Meucciee} framework, and then adapted to 
the CC scattering \cite{Meuccicc} and to the PVES \cite{Meuccipves} . 

Recently, the GFA has been reformulated \cite{eeann} , within a nonrelativistic 
framework, including antisymmetrization and nuclear correlations, that were 
neglected in previous applications. Correlations are included by means of 
realistic one-body density matrices. Their numerical effects on the (e,e$'$) 
reaction are, however, small, within $\sim 5\%$ when only Short-Range 
Correlations (SRC) are included and within $10\%$  when tensor correlations 
are added. These effects are in substantial agreement with those obtained in 
Ref.~\cite{CoL}. 

In the GFA the components of the nuclear response are written in terms of the 
s.p. optical model Green's function. This is the result of suitable 
approximations, such as the assumption of a one-body current and subtler 
approximations related to the IA. 
The explicit calculation of the Green's function can be avoided by its spectral 
representation, which is based on a biorthogonal expansion in terms of a non 
Hermitian optical potential $\cal H$ and of its Hermitian conjugate 
$\cal H^{\dagger}$. In practice, the calculation requires matrix elements of 
the same type as the RDWIA ones in Eq. (\ref{eq.amp}), but involves 
eigenfunctions of both $\cal H$ and $\cal H^{\dagger}$, where the different 
sign of the imaginary part gives in one case an absorption and in the other 
case a gain of flux. Thus, the total flux is conserved and the imaginary part 
is responsible for the redistribution of the strength among different channels. 
This approach guarantees a consistent treatment of FSI in the exclusive and in 
the inclusive scattering.

\begin{figure}[ht]
\begin{center}
\includegraphics[height=70mm,width=70mm]{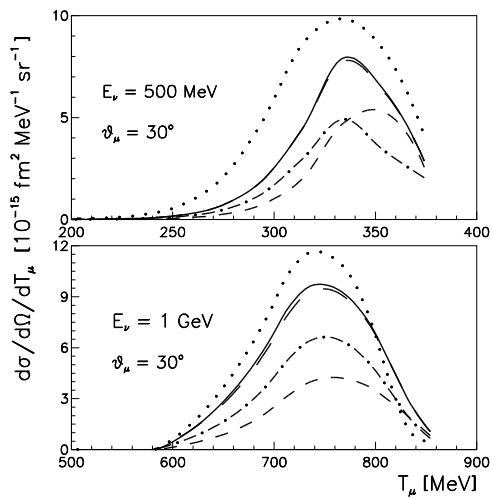}
\vspace{2mm}
\caption{The cross sections of the $^{16}$O$(\nu_{\mu},\mu^-)$ 
reaction for $E_\nu$ = 500 and 1000 MeV at $\theta_\mu = 30^{\mathrm o}$ as a
function of the muon kinetic energy $T_\mu$.  
Results for GFA (solid) RPWIA (dotted), rROP (long-dashed) are compared. 
The dot-dashed lines give the contribution of the integrated exclusive 
reactions with one-nucleon emission. Short dashed lines give the cross sections 
of the $^{16}$O$(\bar\nu_{\mu},\mu^+)$ reaction calculated within the GFA. 
(From Ref.~9).}
\end{center}
\end{figure}
An example of the role of FSI is displayed in Fig.~2, where the 
$^{16}$O$(\nu_{\mu},\mu^-)$  cross sections calculated within the GFA are 
compared with the results of the Relativistic Plane Wave IA (RPWIA), where the 
plane wave approximation is assumed for the outgoing nucleon and FSI are 
neglected. The results obtained when only the real part of the Relativistic 
Optical Potential (rROP) is retained and the imaginary part is neglected are 
also shown in the figure. This approximation conserves the flux, but is 
inconsistent with the exclusive process. Although the use of a complex optical 
potential is conceptually important from a theoretical point of view, the very 
small differences given by the GFA and rROP results mean that the conservation 
of the flux is the most important condition in the present situation. The partial 
contribution given by the sum of all the integrated exclusive one-nucleon 
knockout reactions, also shown in the figure, is much smaller than the complete 
result. The difference is due to the spurious loss of flux produced by the 
absorptive imaginary part of the optical potential. 


For the analysis of data a precise knowledge of $\nu$-nucleus cross sections is 
needed, where theoretical uncertainties are reduced as much as possible. To this 
purpose, it is important to check the differences and the consistencies of the 
different models and the validity of the approximations used.   
 
\begin{figure}[ht]
\begin{center}
\includegraphics[scale=0.45, bb= 80 500 500 750]{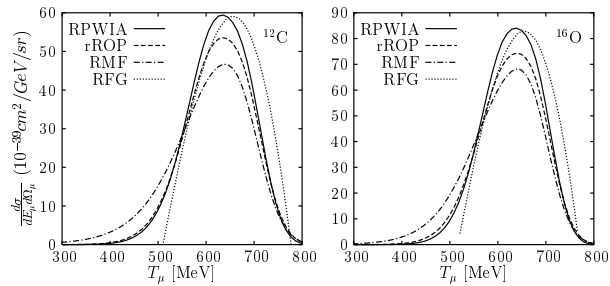}
\vspace{2mm}
\caption{The cross sections of the $^{12}$C$(\nu_{\mu},\mu^-)$ and 
$^{16}$O$(\nu_{\mu},\mu^-)$ reactions for $E_\nu$ = 1000 MeV at 
$\theta_\mu = 45^{\mathrm o}$ as a function of $T_\mu$.  The results of RPWIA 
(solid), rROP (dashed), RMF (dot-dashed), RFG (dotted) are compared. 
(From Ref.~19).}
\end{center}
\end{figure}
Different treatments of FSI are considered in Fig.~3 \cite{Cab} . 
The RPWIA and rROP cross sections are compared with those of a 
Relativistic Mean Field (RMF) approach, where the distorted waves are 
calculated with the same potential used for the initial bound states. 
The results of the Relativistic Fermi Gas (RFG) are also displayed in the 
figure.

CC and NC cross sections calculated in the RPWIA and in the nonrelativistic 
PWIA are compared in Fig.~4. The differences are small at $E_\nu$=500 Mev and
somewhat  larger at 1 GeV.
\begin{figure}[ht]
\begin{center}
\includegraphics[height=90mm,width=90mm]{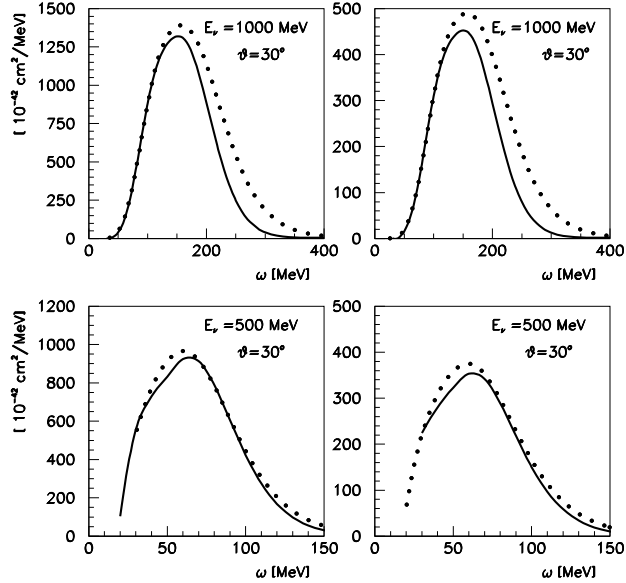}
\vspace{2mm}
\caption{The cross sections of the  $^{16}$O$(\nu_e,e^-)$ (left panel) and 
$^{16}$O$(\nu,\nu')$ (right panel) reactions for $E_\nu$ = 500 and 1000 MeV at
$\theta = 30^{\mathrm o}$. Solid lines: RWPIA (A. Meucci) 
dotted lines non relativistic PWIA (G. Co').} 
\end{center}
\end{figure}

Relativistic and nonrelativistic models have been developed and 
applied in nuclear physics with electroweak probes. Relativistic 
effects have been widely investigated. They increase with the energy and for 
energies of about 1 GeV a fully relativistic model should be used. 
In order to account for these effects, relativistic corrections in the 
kinematics and in the current operators are often included in 
nonrelativistic models. Even with these corrections, however, a 
nonrelativistic approach cannot reproduce  all the relativistic aspects in the 
dynamics of a relativistic one. Moreover, calculations are generally carried 
out in the two cases with different theoretical ingredients. Relativistic and 
 nonrelativistic models have thus to be considered as different and alternative 
 approaches. All the available models make use of approximations and have 
 merits and shortcomings. Only a relativistic approach can fully account for 
relativistic effects. At present, however, nonrelativistic models may allow to 
include specific nuclear effects, e.g., due to correlations and two-body 
currents, in a more consistent and clear framework.  

Nuclear correlations and FSI in electron and neutrino scattering off $^{16}$O 
are considered in Refs.~\cite{Benhar,BenharN,Benhar1}. The approach is based 
on the nonrelativistic nuclear many-body theory and on the IA. In the IA the 
scattering process off a nuclear target reduces to the incoherent sum of 
elementary processes involving only one nucleon. For each term in the sum the cross section is factorized into the 
product of the elementary lepton-nucleon cross section and the nuclear 
spectral function, describing the momentum and energy distribution of nucleons 
in the target. The correlated spectral function of $^{16}$O is obtained with a 
local density approximation in which nuclear matter results for a wide range 
of density values are combined with the experimental information from the
$^{16}$O(e,e$'$p) knockout reaction. FSI are treated within a Correlated 
Glauber Approximation (CGA), which rests on the premises that: i) the struck 
nucleon moves along a straight trajectory with constant velocity (eikonal 
approximation), ii) the spectator nucleons are seen by the struck particle 
as a collection of fixed scattering centers (frozen approximation). 
Under these assumptions, the propagator of the struck nucleon after the 
electroweak interaction is factorized in terms of the free space propagator and
of a part that is related to the nuclear transparency measured in (e,e$'$p).
The cross section is written in the convolution form in terms of the IA
cross section and of a folding function including FSI. 
\begin{figure}[ht]
\begin{center}
\includegraphics[height=50mm,width=80mm]{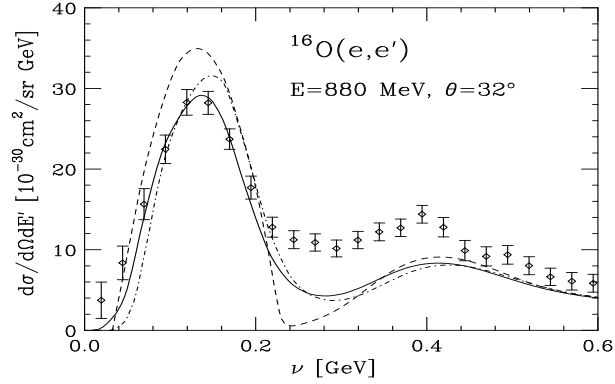}
\vspace{2mm}
\caption{Cross section of the $^{16}$O(e,e$'$) reaction at beam energy 880 MeV 
and electron scattering angle 32$^\circ$. Dot-dashed line: IA, solid line: full
calculation including FSI, dashed line: FG. Data from Ref.~7. 
(From Ref.~20).}
\end{center}
\end{figure}
A numerical example is presented in Fig.~5 for the  $^{16}$O(e,e$'$) reaction. 
The model gives a good description of data in the region of the QE peak. FSI 
produce a shift and a redistribution of the strenght leading to a quenching of 
the peak and to an enhancement of the tail. The FG model overestimates the 
data. The failure of the calculations to reproduce the data in the 
$\Delta$-peak region is likely to be mostly due \cite{BenharM} to the poor 
knowledge of the neutron structure function at low $Q^2$.
The ability to yield quantitative predictions over a wide range of 
energies is critical to the analysis of neutrino experiments, in which the 
energy of the incident neutrino is not known, and must be reconstructed from 
the kinematics of the outgoing lepton. 

\section{Two-Body Weak Axial Current}

Two-body Weak Axial Exchange Currents (WAECs) are considered in 
Refs.~\cite{MR,MR1}. The issue of the pion pair axial current is
addressed in Ref.~\cite{MR}, showing how the interplay of the chiral 
invariance and the double-counting problem restricts uniquely its form. 
The complete treatment of the WAECs, including the heavy meson exchange 
contributions, that is, besides $\pi$-, $\rho$-$\omega$-$a_1$-exchange, is the 
main result of Ref.~\cite{MR1}.

The Partial Conservation of the Axial Current (PCAC) reads
\begin{equation}
q_\mu\,<\Psi_f|j^a_{5\mu}(q)|\Psi_i>\,=\,if_\pi
 m^2_\pi\Delta^\pi_F(q^2)\, <\Psi_f|m^a_\pi(q)|\Psi_i>\,,
\label{eq.pcac}
\end{equation}   
where the weak axial  current $j^a_{5\mu}(q)$ is the sum of the one- and 
two--nucleon components
\begin{equation}
j^a_{5\mu}(q)\,=\,\sum^A_{i=1}\,j^{a}_{5\mu}(1,i,q_i)\,+\,\sum^A_{i<j}\,
j^{a}_{5\mu}(2,ij,q) ,
\label{eq.cur}
\end{equation} 
$m^a_\pi(q)$ is the pion source (the pion production/absorption amplitude).  
The wave functions $|\Psi_{i,f}>$ describe the initial or final nuclear states, 
which are eigenfunctions of the Schr\"odinger equation with the nuclear 
Hamiltonian $H=T+V$, where $V$ is the potential describing the interaction
between nucleon pairs. From Eqs. (\ref{eq.pcac}) and (\ref{eq.cur}), taking for 
simplicity $A=2$, the following set of equations are obtained in the operator 
form for the one- and two-nucleon components of the axial current
\begin{eqnarray}
\vec q_i \cdot \vec j^a_{\,5}(1,\vec
q_i)\,&=&\,[\,T_i\,,\,\rho^a_{\,5} (1,\vec q_i)\,]\,+\,if_\pi
m^2_\pi \Delta^\pi_F(q^2) m^a_\pi(1,\vec q_i)
\,,\quad i=1,2\,,    \\
\vec q \cdot \vec j^a_{\,5}(2,\vec q)\,&=&\,[\,T_1+T_2\,,\,
\rho^a_{\,5}(2,\vec q)\,]\,+\,([\,V\,,\,\rho^a_{\,5}(1,\vec q)\,] 
+{(1\leftrightarrow 2)} \nonumber \\
& & +if_\pi m^2_\pi \Delta^\pi_F(q^2)m^a_\pi(2,\vec q).
\label{eq.ccc}
\end{eqnarray}
If the WAECs are constructed in order to satisfy these conditions, the matrix 
element of the total current between the solutions of the nuclear 
equation of motion satisfies the PCAC.

The WAECs are constructed from an effective Lagrangian possessing the chiral 
symmetry and respecting the vector dominance model. The exchange amplitudes
of range $B$ ($B$=$\pi$, $\rho$, $\omega$, $a_1$) are derived as Feynman tree 
graphs and satisfy the PCAC. The currents are constructed from the amplitudes, 
in analogy with the electromagnetic Meson Exchange Currents (MEC) \cite{ATA}~, 
as the difference between the relativistic amplitudes and the first Born 
iteration of the weak axial one--nucleon current contribution to the 
two-nucleon scattering amplitude, satisfying the Lippmann-Schwinger equation. 

For practical calculations a nonrelativistic reduction of the currents is 
performed. The nuclear wave functions are generated by the same One-Boson 
Exchange Potential (OBEP) of Eq. (\ref{eq.ccc}) and the same 
couplings and strong form factors as in the potential are applied in
the WAECs. Consistent calculations are carried out  employing the realistic 
OBE potentials OBEPQG \cite{OPT} , Nijmegen 93 and Nijmegen I \cite{SKTS} .

The WAECs have been used to calculate $\nu$- and $\bar\nu$-deuteron 
disintegration cross sections at the typical solar neutrino energies. 
The results indicate that the main two-body effect comes from the $\Delta$ 
excitation and that the heavy-meson exchange contributions are of the same 
order of magnitude as the $\pi$-exchange one. The uncertainty of standard
nuclear physics calculations has been reduced from 5-10\% to about 3\%.

\section{Scaling and Superscaling in Lepton-Nucleus Scattering}

The analyses of ongoing and future neutrino experiments require reliable 
predictions of $\nu$-nucleus cross sections.
Any nuclear model should first be tested in comparison with electron scattering 
data. Sophisticated models have been developed to describe
electron-nucleus scattering. In spite of all these efforts, however, the 
uncertainty due to the treatment of nuclear effects in different models is still 
high when compared with the required precision.
 
The analogies between $\nu$-nucleus and electron-nucleus scattering suggest 
an alternative approach to extract model independent $\nu$-nucleus cross 
sections from experimental electron-nucleus cross sections \cite{scal} . 
Instead of using a specific model, one can exploit the scaling properties of
(e,e$'$) data and i) extract a scaling function from (e,e$'$) data, ii) invert
the procedure to predict CC  $\nu$-nucleus cross sections. 

This scaling approach \cite{Cab,scal,scalsr,scalnc,scalsupp} relies on the 
superscaling properties of the electron scattering data \cite{scaling} . At 
sufficiently high momentum transfer a scaling function is derived dividing the 
experimental (e,e$'$) cross sections by an appropriate single-nucleon cross 
section. This is basically the idea of the IA. If this scaling function depends 
only upon one kinematical variable, the scaling variable, one has scaling of 
first kind. If the scaling function is roughly the same for all nuclei, one has 
scaling of second kind. When both kinds of scaling are fulfilled, one says that 
superscaling occurs. 

In the QE region the scaling variable is \cite{scal}
\begin{equation}
\psi_{\rm QE}=\pm\sqrt{1/(2T_F)\left(q\sqrt{1+1/\tau} -\omega -1\right)},
\end{equation}
where $T_F$ is the Fermi kinetic energy, $4m_{\mathrm N}^2\tau=q^2-\omega^2$
and the $-$ ($+$) sign corresponds to energy transfers lower (higher) than the
QE-peak ($\psi=0$).

An extensive analysis of electron scattering data \cite{scaling} has shown 
that scaling of first kind is fulfilled at the left of the QEP and broken 
at its right, whereas scaling of second kind is well satisfied at the
left of the peak and not so badly violated at its right. As a consequence,  
a scaling function $f^{\mathrm{QE}}$ can be extracted from the data.

The superscaling analysis has been extended to the first resonance peak 
\cite{scal} . The contribution of the $\Delta$ has been (approximately) 
isolated in the data by subtracting the QE scaling contribution 
from the total experimental cross sections. Then, the scaling function has been 
studied as a function of a new scaling variable
\begin{equation}
\psi_\Delta=\pm\sqrt{1/(2T_F)\left(q\sqrt{\rho+1/\tau} -\omega\rho -1\right)}~,
\end{equation}
where $\rho=1+(m_\Delta^2-m_{\mathrm N}^2)/(4\tau m_{\mathrm N}^2)$.
The results show that also in this region superscaling is satisfied  
and a second superscaling function, $f^\Delta$, can be extracted from the data 
to account for the nuclear dynamics.
Clearly this approach can work only at $\psi_\Delta < 0$, since at 
$\psi_\Delta > 0$ other resonances and the tail of the deep-inelastic 
scattering start contributing.

The two scaling functions can firstly be tested in comparison with (e,e$'$) 
data and can then be used to predict 
$\nu$-nucleus cross sections.

\begin{figure}[ht]
\begin{center}
\includegraphics[scale=0.4]{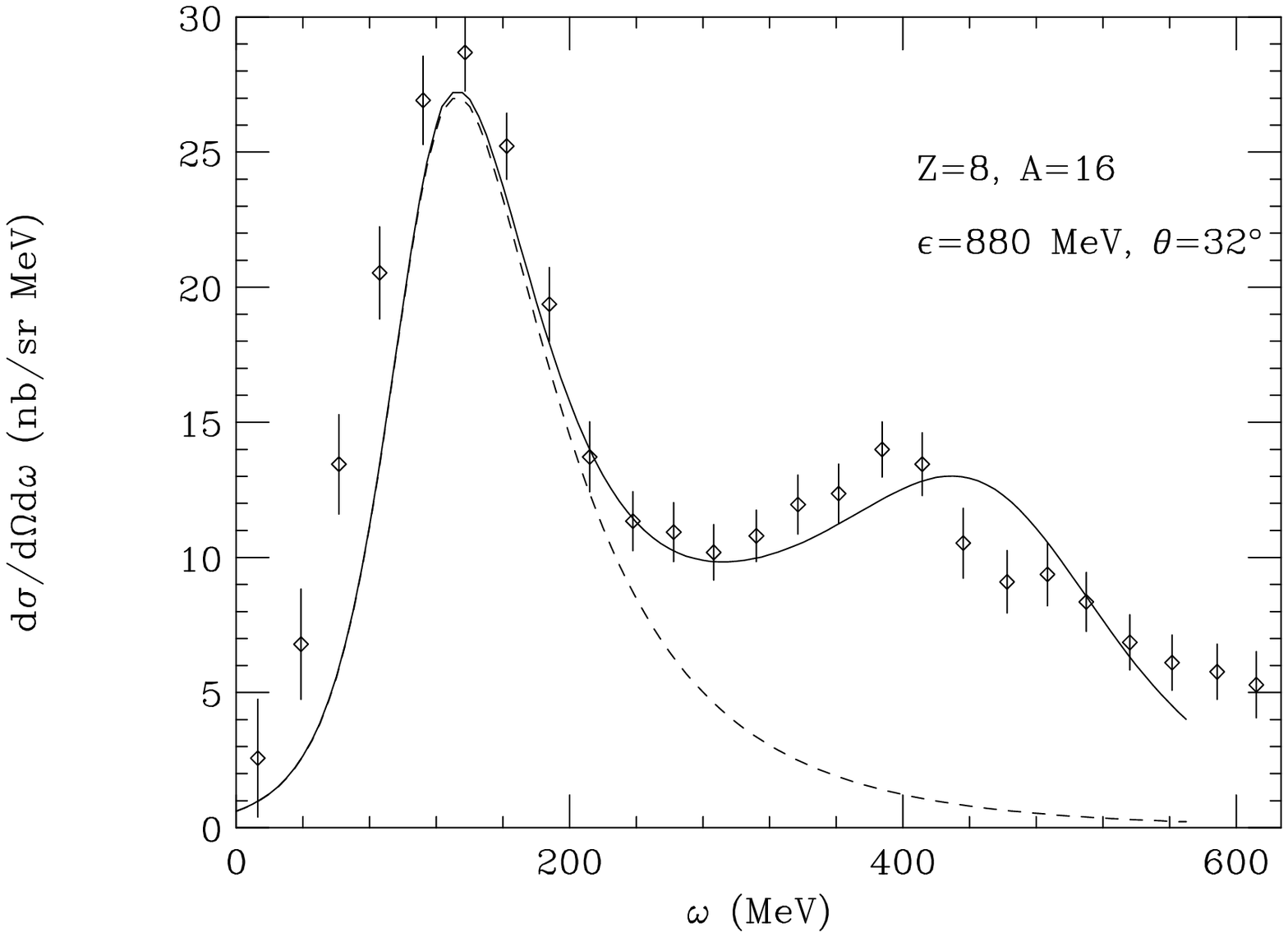}
\includegraphics[scale=0.4]{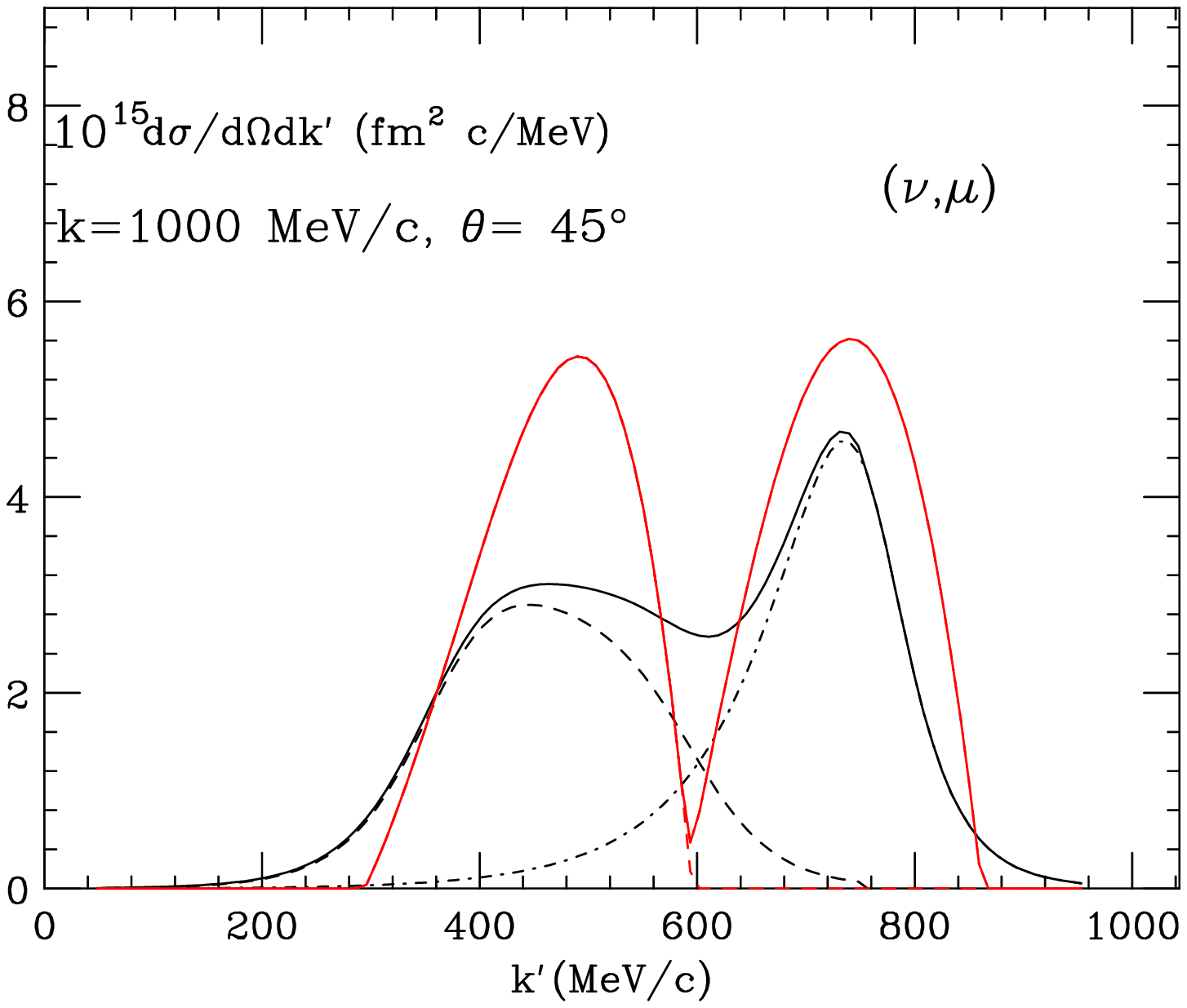}
\vspace{2mm}
\caption{Left panel: cross section of the $^{16}$O(e,e$'$) reaction. The solid 
line is obtained using $f^{\mathrm{QE}}$ and $f^\Delta$, the dashed line
using only $f^{\mathrm{QE}}$. Data from Ref.~7. 
Right panel: cross section of the $^{12}$C($\nu,\mu^-$) reaction as
a function of the muon momentum. Solid line: superscaling prediction; heavier 
line: RFG. The separate  QE and $\Delta$ contribution are shown 
(dotted lines). (From Ref.~29).}
\end{center}
\end{figure}
In the left panel of Fig.~6  an example of the (e,e$'$) cross section 
reconstructed by multiplying the empirical superscaling functions by the 
appropriate single-nucleon functions is shown in comparison with data. In this 
as well as in many other cases it turns out that typical deviations  are 10\% 
or less \cite{scal} , thus confirming that the scaling approach offers a 
reliable description of the nuclear dynamics.

A numerical prediction for CC $\nu$-scattering is displayed in the right panel 
of Fig.~6. The result obtained using the empirical scaling functions 
$f^{\mathrm{QE}}$ and $f^\Delta$ is compared with the result of 
the RFG model. The RFG cross section differs significantly from the scaling 
prediction, which lies somewhat lower and extends over a wider range in $k'$.

In the case of NC reactions, the kinematics are different from the ones of CC 
and (e,e$'$) processes since the detected final state is the outgoing nucleon 
and the neutrino kinematic variables are integrated over. This implies
an integration region in the residual nucleus variables which is different in 
the two cases and it is not obvious that the superscaling procedure, based
on the analogy with inclusive electron scattering, is still valid. 
The scaling method is based on a factorization assumption which has to be 
tested numerically. The outcome is that the procedure can be applied also 
to NC reactions \cite{scalnc} .

The properties of the empirical scaling functions should be accounted for by
microscopic calculations. In particular, the asymmetric shape of $f^{\mathrm{QE}}$
should be explained. In Ref.~\cite{Cab} the scaling properties of 
different models (RPWIA, rROP, RMF) are verified. Superscaling is fulfilled to 
high accuracy in the QE region by the three descriptions of FSI 
considered. Then, the associated scaling functions are compared with the 
experimental scaling function. Only the RMF model is able to reproduce the  
asymmetric shape of the experimental function. This result deserves further 
investigation.

More results on the scaling approach can be found in \cite{Martini} .

\section{Electromagnetic Reactions on Few-Nucleons Systems}   

Electromagnetic reactions on few-body nuclei are investigated in
Refs.~\cite{few1,few2,few3}. The theoretical study of the electromagnetic 
structure of few-body nuclei requires the knowledge of the nuclear wave 
functions and the electromagnetic transition operators. 
For the low-energy observables considered in Ref.~\cite{few1} and for 
processes involving two and three nucleons, accurate bound 
and scattering states are calculated using the pair-correlated 
Hyperspherical Harmonics (HH) method \cite{Kieetal} from the Argonne $v_{18}$ 
(AV18) two-nucleon~\cite{Wir95} and Urbana IX  (UIX) \cite{Pud95} or 
Tucson-Melbourne (TM) \cite{Coo79} three-nucleon interactions. 
For two- and three-nucleon interactions the nuclear electromagnetic current 
operator includes, in addition to the one-body terms, also two- and three-body 
terms. Different models for conserved two- and three-body currents are 
constructed using either meson exchange (ME) mechanisms or minimal substitution 
(MS) in the momentum dependence of the interactions. The connection between 
these two schemes is elucidated \cite{few1} .

The electromagnetic current operator must satisfy the Current Conservation 
Relation (CCR)
\begin{equation}
  {\bf q}\cdot{\bf j}({\bf q})= [H,\rho({\bf q})]\ ,\label{eq.ccr}
\end{equation}
where the Hamiltonian $H$ contains two- and three-body interactions, 
$v_{ij}$ and $V_{ijk}$, respectively.
To lowest order in $1/m$, Eq.~(\ref{eq.ccr}) separates into
\begin{equation}
  {\bf q}\cdot{\bf j}_i({\bf q}) = 
\biggl[{\frac{{\bf p}_i^2}{2m}},\rho_i({\bf q})\biggr]  \ ,\hspace{0.5cm}
  {\bf q}\cdot{\bf j}_{ij}({\bf q}) = [v_{ij},\rho_i({\bf q})+\rho_j({\bf q})]
 \ , \label{eq.ccr2}
\end{equation}
and similarly for the three-body current. The one-body current satisfies the 
CCR. It is rather difficult to construct conserved two- and three-body currents
because  $H$ includes momentum and isospin dependent terms that do not commute
with $\rho$.

The two-body current can be separated into model-independent (MI) and 
model-dependent (MD) parts. The MD current is purely transverse and is 
unconstrained by the CCR.  
The MI current has longitudinal and transverse components and is constrained 
by the CCR. The longitudinal part is constructed so as to satisfy the CCR. 
The MI currents from the momentum-independent terms of  AV18 have been 
constructed following the ME scheme and
satisfy the CCR. The currents from the momentum-dependent terms of the 
interaction obtained in the ME scheme do not strictly satisfy the CCR. If these 
currents are obtained in the MS scheme they satisfy the CCR. 
Both the ME and MS schemes can be generalized to calculate the three-body 
currents induced by a three-nucleon interaction.

Several electronuclear observables have been calculated \cite{few1} to 
test the model of the current operator. For the $A$=3 
nuclear systems, cross sections and polarization observables in the energy 
range 0--20 MeV are compared with data and with  earlier results 
\cite{Viv00}  where the current operator retains 
only two-body terms, all of them obtained within the ME scheme, and the CCR is 
only approximately satisfied.  

The differences with respect to the previous results  
are generally small, but for some of the polarization parameters measured in 
pd radiative capture, specifically the tensor polarizations $T_{20}$ and 
$T_{21}$, where the exactly conserved currents resolve the 
discrepancies between theory and data obtained in Ref.~\cite{Viv00}. An 
example is shown in Fig.~7. 
\vspace{10mm}
\begin{figure}[ht]
\begin{center}
\includegraphics[height=50mm,width=90mm]{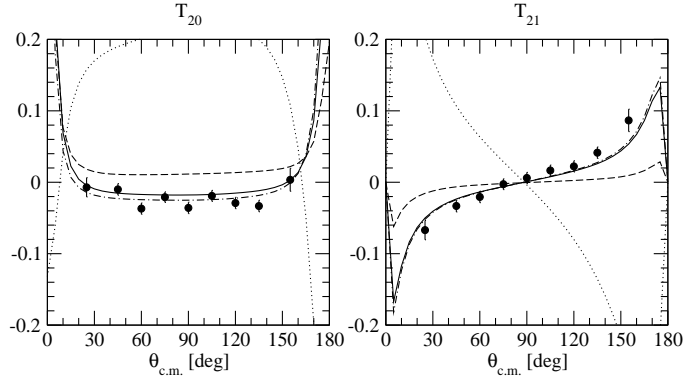}
\vspace{2mm}
\caption{Deuteron tensor polarization observables $T_{20}$ and $T_{21}$ 
for pd radiative capture at $E_{c.m.}$= 2 Me. 
The dotted curves include only the one-body current, the dashed 
curves are the results of Ref.~43, the dot-dashed curves are 
obtained in the long-wavelength-approximation (LWA), applying the Siegert 
theorem. The solid curves are the results of Ref.~35. 
Data from Ref.~42. (From Ref.~35).} 
\end{center}
\end{figure}

An overall nice description has been reached for all the observables. Also, 
some small three-body currents effects are noticeable, which is an indication 
of the fact that if a Hamiltonian model with two- and three-nucleon 
interactions is used, then the model for the nuclear current operator should 
include the corresponding two- and three-body contributions \cite{few1} .

A comparison with the $^3$He(e,e$'$p)d 
\cite{Rvachev05} and $^4$He($\vec{\mathrm e}$,e$'\vec{\mathrm p}$)$^3$H 
\cite{Strauch03} JLab data is presented in Refs.~\cite{few2} and 
\cite{few3}, respectively. 
Accurate bound state wave functions are still calculated within the HH method.
The electromagnetic current operator includes one- and two-body terms. Since 
at the high energies of the JLab experiments a theory of the interaction and 
few-nucleon systems is not available, different treatments of FSI are 
used. In Ref.~\cite{few2}, a Glauber approximation is used, where the 
profile operator in the Glauber expansion is derived from a NN scattering 
amplitude, which retains its full spin and isospin dependence, and is 
consistent with phaseshift analyses of NN scattering data. 
In Ref.~\cite{few3}, a nonrelativistic optical potential is used including 
charge-exchange terms. In both cases a fair agreement with data is found. 
Of particular interest is the agreement obtained \cite{few3} for the ratio of 
transverse to longitudinal polarization transfers in the 
$^4$He($\vec{\mathrm e}$,e$'\vec{\mathrm p}$)$^3$H reaction. In the elastic 
process $\vec{\mathrm e}{\mathrm p}\rightarrow {\mathrm e}\vec{\mathrm p}$, 
the ratio is proportional to that of the electric to magnetic form factors of 
the proton, and a measurement in a nucleus by QE proton knockout can shed 
light, in principle, on the question of whether these form factors are 
modified in medium. Thus, the agreement found when free-nucleon electromagnetic 
form factors are used in the current operator \cite{few3} challenges the current 
interpretation of data in terms of medium-modified form factors.

\section{Electromagnetic Reactions on Complex Nuclei}

Complementary polarization measurements suited to study nucleon properties in 
the nuclear medium are proposed in Ref.~\cite{skew}, where the general 
formalism of the 
$\vec{\mathrm A}$($\vec{\mathrm e},{\mathrm e}'\vec{\mathrm p}$)B 
reaction is presented  within the RPWIA.  The simultaneous polarization of the 
target and the ejected proton provides information which 
is not contained in the  $\vec{\mathrm A}$($\vec{\mathrm e}$,e$'$p)B and 
A($\vec{\mathrm e}$,e$'\vec{\mathrm p}$)B reactions.  The 
polarization transfer mechanism in which the electron interacts with the 
initial nucleon carrying the target polarization, making the proton exit
with a fractional polarization in a different direction, is referred
to as ``skewed polarization''. Although difficult to measure, these new 
observables would provide information on nucleon properties complementing 
the results for the ratio of transverse to longitudinal polarization 
transfers.

Proton emission induced by polarized photons of energies  above the 
giant resonance region and below the pion production threshold is studied in 
Ref.~\cite{gp}.
With respect to (e,e$'$p), a different kinematics is explored in the
($\gamma$,p) reaction. In fact, for a real photon the energy and 
momentum transfer are constrained by the condition $\omega=|\q|$, and 
only the high-momentum components of the nuclear wave function are probed.
Moreover, the validity of the direct knockout mechanism, which is clearly 
stated for (e,e$'$p), is questionable for ($\gamma$,p), where important 
contributions are due to two-nucleon processes, such as those involving MEC 
\cite{MEC,MECCO} .  
The sensitivity of various polarization observables in the ($\vec{\gamma}$,p) 
reaction to FSI, MEC, and SRC is investigated \cite{gp} using the same model 
\cite{MECCO} applied to calculate ($\gamma$,p) cross sections. 

The sensitivity to FSI, MEC, and NN correlations, in particular SRC, are the 
same issues considered in the theoretical studies of electron- and 
photon-induced two-nucleon knockout \cite{CO2N,TNKO} . Since a long time these 
reactions have been devised as a preferential tool to investigate SRC \cite{book} . 
In fact, direct insight into SRC can be obtained from the situation where the 
electromagnetic probe hits, through a one-body current, either nucleon of a 
correlated pair and both nucleons are then ejected from the nucleus. This 
process is entirely due to correlations. Additional complications have, however, 
to be taken into account, such as competing mechanisms, like  contributions of 
two-body MEC and $\Delta$ isobar excitations, as well as the FSI between the 
two outgoing nucleons and the residual nucleus. 

The calculated cross sections are sensitive to the different ingredients of the
model and to their treatment. The role and relevance of competing reaction 
mechanisms and of different contributions is different in different reactions 
and kinematics. It is thus possible, in principle, with the help of 
theoretical predictions, to envisage appropriate situations where specific effects can be 
disentangled and separately investigated. 
Data from NIKHEF \cite{NIKHEF} and MAMI  \cite{MAMI} for the exclusive 
$^{16}$O(e,e$'$pp)$^{14}$C reaction confirmed the validity of the direct 
knockout mechanism and gave clear evidence of SRC for the transition to the  
ground state of $^{14}$C. This important result, that was obtained from a close
collaboration between experimentalists and theorists,  means that further
studies on these reactions would make it possible to disentagle SRC in 
experimental cross sections. From the experimental side, however, during the 
last few years only the results of a first measurement of the  
$^{16}$O(e,e$'$pn)$^{14}$N reaction have been published \cite{pn} . 
From the theoretical side, the recent studies ~\cite{CO2N,TNKO}
focussed on specific aspects of the theoretical models, such as a consistent 
treatment of different types of correlations in the two-nucleon wave function, 
the competing contribution of correlations and two-body currents, the 
uncertainties in the treatment of the $\Delta$-current, the effects of FSI, 
whose consistent treatment requires in general a genuine three-body approach 
for the mutual interaction of the two nucleons and the residual nucleus.

The outcome of this work is that electromagnetic two-nucleon knockout 
contains a wealth of information on correlations and on the behavior of the 
$\Delta$-current in a nucleus. The uncertainties in the treatment of the 
theoretical ingredients and the large number of parameters involved in the 
models make it difficult to extract clear and unambiguous information  
from one ideal kinematics. Data are therefore needed for electron- and 
photon-induced pp and pn emission and in various kinematics which mutually 
supplement each other. The choice of suitable conditions for the experiments 
and the interpretation of data require close collaboration between 
experimentalists and theorists.

\section*{}

I dedicate this report to the memory of Adelchi Fabrocini, who gave
significant contributions to the study of correlations in electromagnetic
reactions and with whom, during the years, I had so many fruitful and pleasant 
conversations. I thank Franco Pacati for his valuable help and Giampaolo Co' 
for useful discussions.

\end{document}